\documentclass[12pt]{article}
\usepackage{psfig}
\newcommand{\gsim}{\lower.7ex\hbox{$
    \;\stackrel{\textstyle>}{\sim}\;$}}
\newcommand{\lsim}{\lower.7ex\hbox{$
    \;\stackrel{\textstyle<}{\sim}\;$}}
\def\order#1{{\cal O}\left(#1\right)}
\def\eq#1{(\ref{#1})}

\newcommand{\ba}{\begin{eqnarray}}
\newcommand{\ea}{\end{eqnarray}}

\begin{document}

\setlength{\unitlength}{1mm}

\thispagestyle{empty}


\begin{flushright}
{\bf Alberta Thy 06'-01}\\
{\bf hep-ph/0106237}\\[2mm]
{\bf March 2002}
\end{flushright}
\vspace{0.6cm}
\boldmath
\begin{center}
\Large\bf\boldmath Electromagnetic suppression of the decay 
 $\mu\to e\gamma$ 
\end{center}
\unboldmath
\vspace{0.8cm}

\begin{center}

{\large Andrzej Czarnecki 
 }\\[2mm]
{\sl Department of Physics, University of Alberta\\
Edmonton, AB, Canada T6G 2J1}\\[8mm]

{\large and} \\[6mm]

{\large Ernest Jankowski}\\[2mm]
{\sl Department of Physics, University of Alberta\\
Edmonton, AB, Canada T6G 2J1}\\[2mm]
{\sl and}\\[2mm]
{\sl CERN, PPE\\
CH-1211 Geneva 23, Switzerland}

\end{center}

\vfill

\begin{abstract}
Due to large QED anomalous dimensions of the electric and magnetic
dipole operators, the rate of the rare muon decay $\mu\to e\gamma$ is
suppressed by the factor $\left(1-{8\alpha\over \pi}\ln{\Lambda\over
m_\mu}\right)$, independent of the physics responsible for the
lepton-flavor violation, except for the scale $\Lambda$ at which it
occurs. For $\Lambda =100\ldots 1000$ GeV, the resulting decrease of
the rate amounts to about $12\ldots 17$\%.
\end{abstract}
\vfill

\newpage

\section{Introduction}

\begin{figure}[htb]
\hspace*{15mm}
\begin{minipage}{16.cm}
\vspace*{5mm}
\begin{tabular}{c@{\hspace{15mm}}c}
\psfig{figure=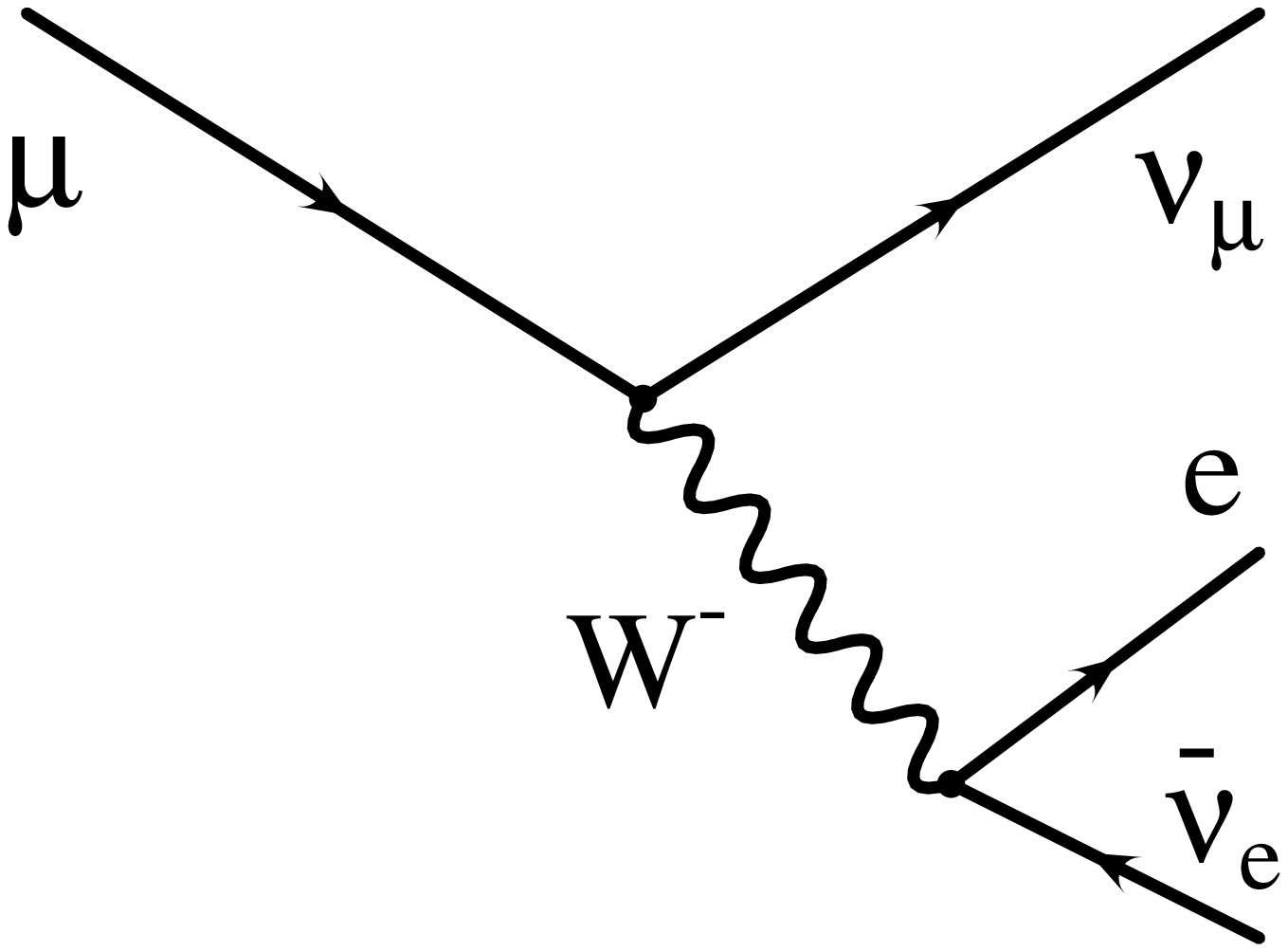,width=40mm}
&
\psfig{figure=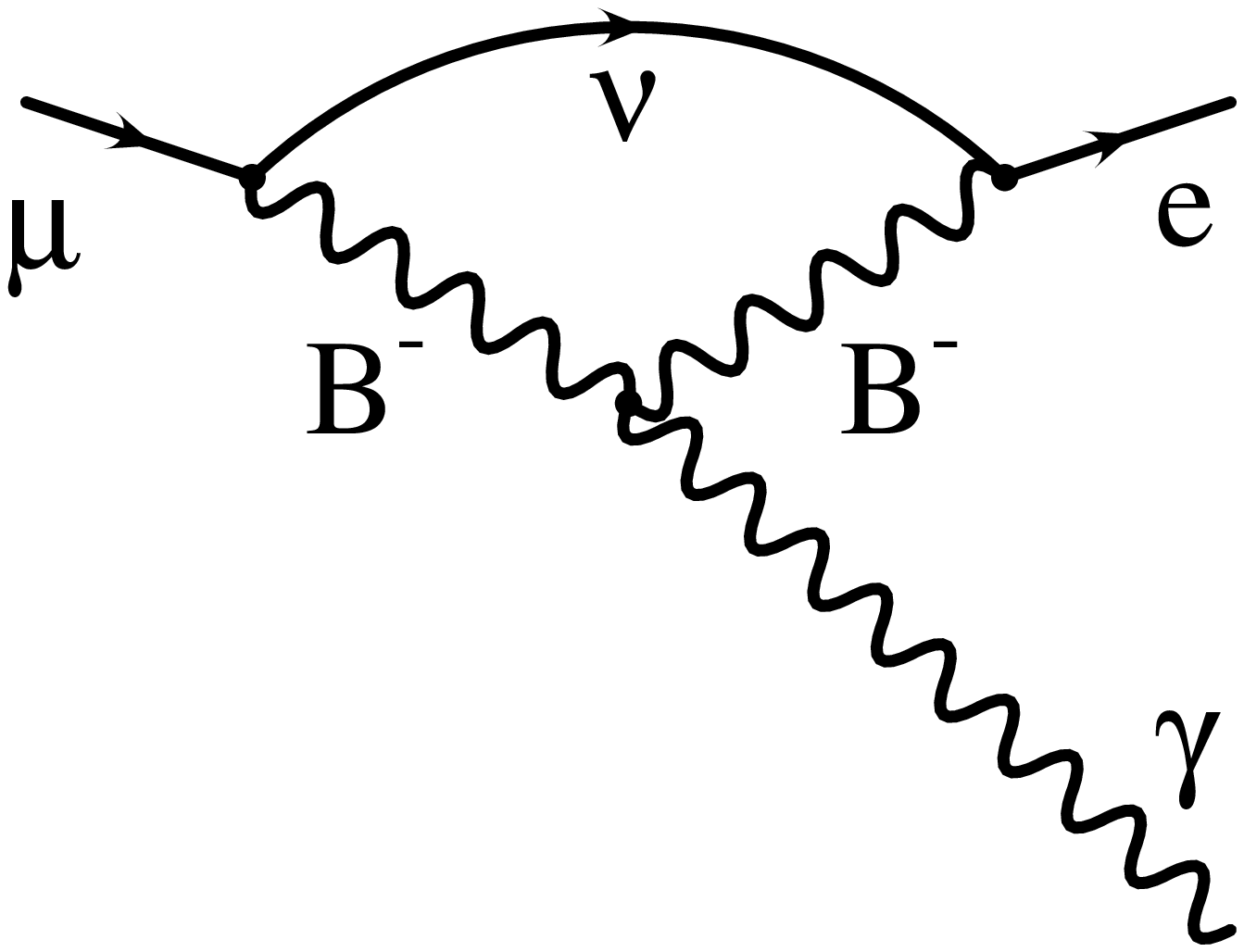,width=40mm}\\
(a) & (b)
\end{tabular}
\end{minipage}
\caption{\sf (a) Ordinary muon decay; (b) The puzzle of $\mu\to
e\gamma$ absence in the early models with an intermediate vector boson.}
\label{fig1}
\end{figure}

The only observed decay channel of the muon is $\mu^- \to e^-
\bar\nu_e\nu_\mu$ (with possible photon or electron-positron pair
emission).  However, since the discovery of the muon more than half
century ago, searches have been undertaken for the decay $\mu\to
e\gamma$.  Initially, when the muon was thought to be an excited state
of the electron, this was expected to be its dominant decay channel.
It was soon realized that it is very strongly suppressed (the
early experiments are summarized in \cite{Crittenden61}).  When an
intermediate boson was proposed to explain the mechanism of weak
interactions \cite{Yukawa49}, the absence of $\mu\to e\gamma$ led to
the hypothesis that the two neutrinos in the muon decay
(Fig.~\ref{fig1}(a)) have different flavors so that the interaction shown
in Fig.~\ref{fig1}(b) cannot occur \cite{Lee60,LeeYang60}.  
The
existence of the muon neutrino, distinct from the electron one, was
demonstrated in the classic 1962 experiment in Brookhaven
\cite{Danby62}.   In this way, the limits placed on the branching
ratio for $\mu\to e\gamma$ helped establish the concept of families or
generations of fermions, which became one of the cornerstones of the
standard model.

\begin{figure}[htb]
\hspace*{15mm}
\begin{minipage}{16.cm}
\vspace*{1mm}
\begin{tabular}{c@{\hspace{15mm}}c}
\psfig{figure=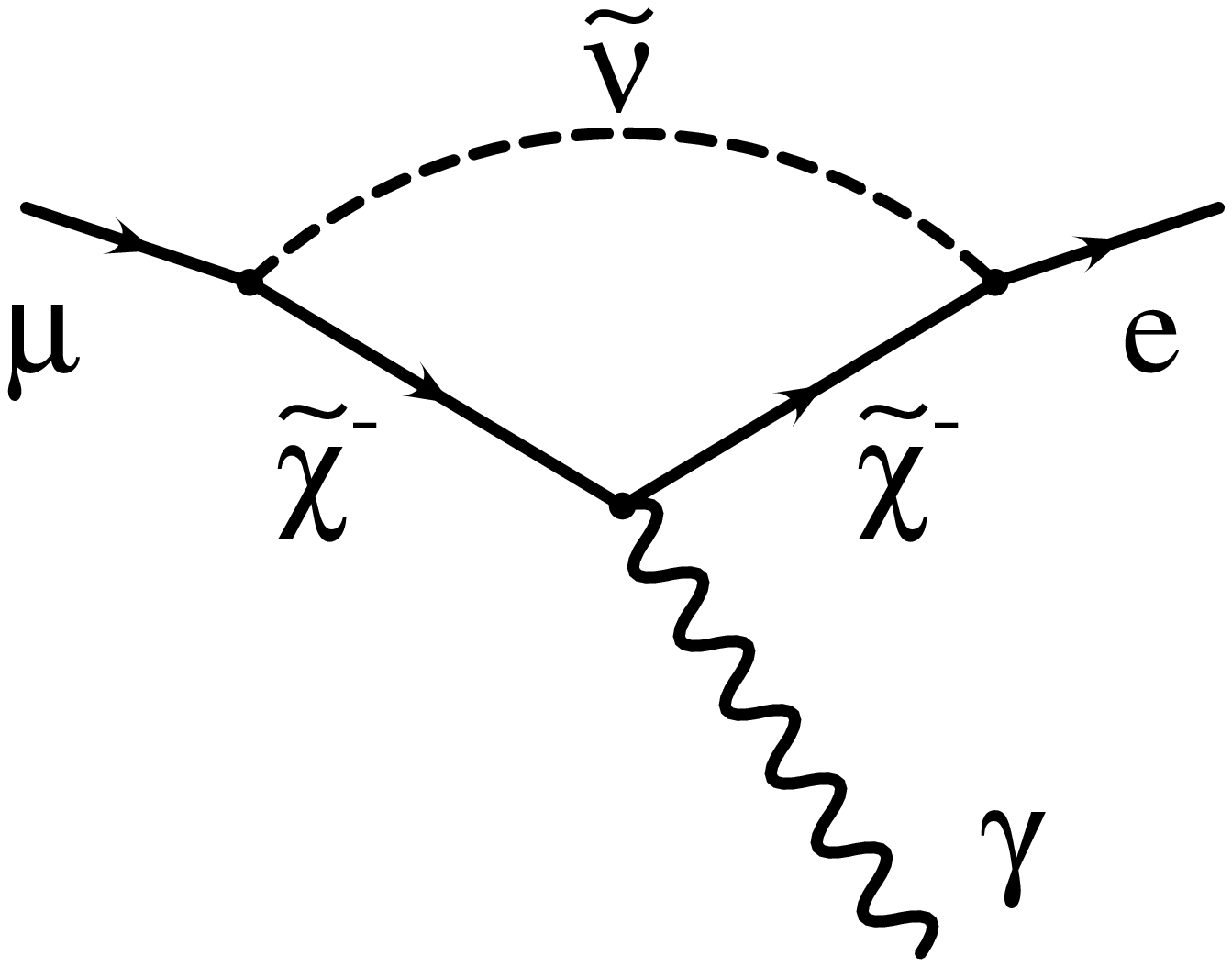,width=40mm}
&
\psfig{figure=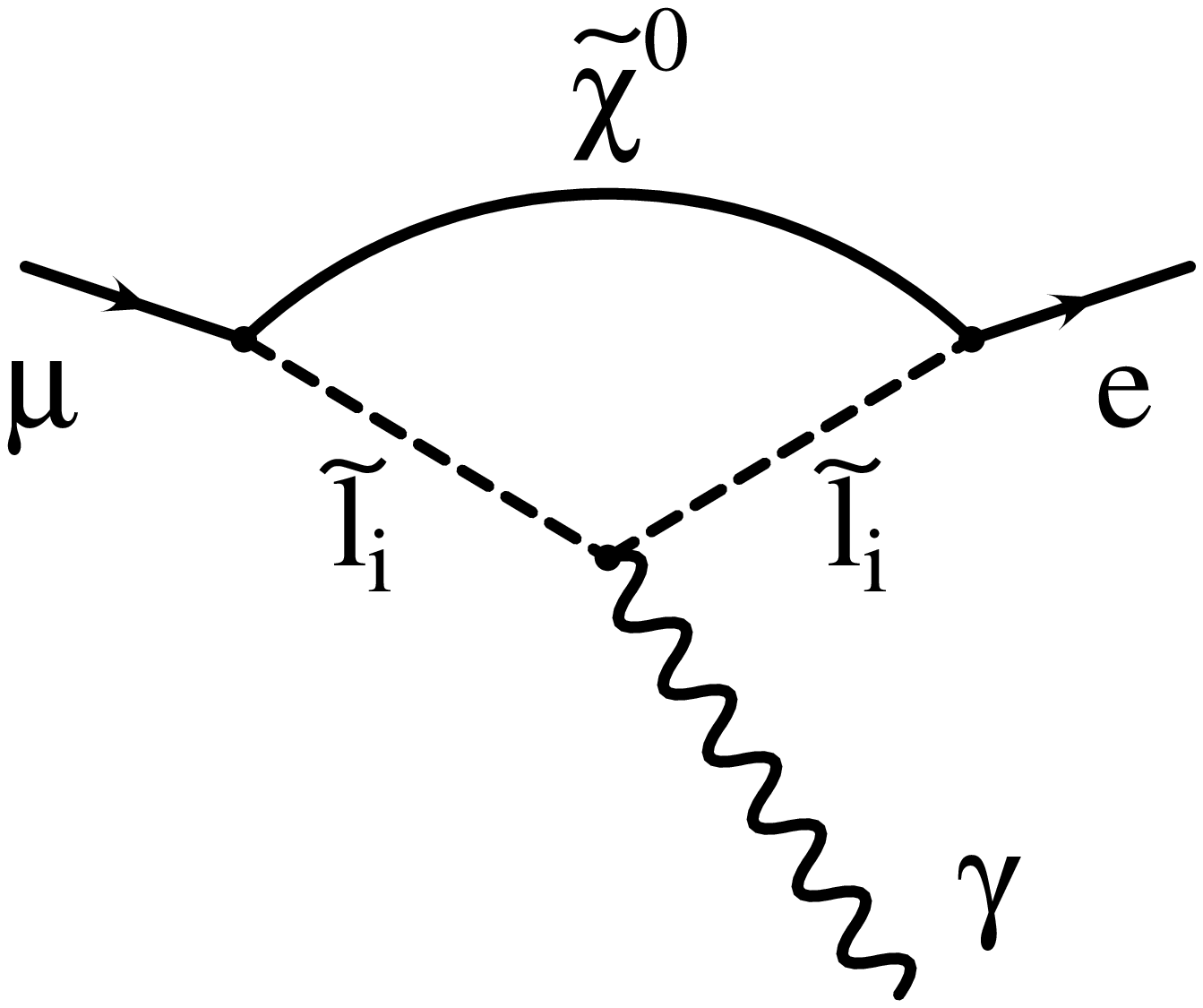,width=40mm}\\
(a) & (b)
\end{tabular}
\end{minipage}
\caption{\sf Supersymmetric amplitudes which might give rise to the decay
$\mu\to e\gamma$.} 
\label{fig2}
\end{figure}

In fact, the standard model with massless neutrinos strictly forbids
the lepton-flavor nonconserving transitions like $\mu\to e\gamma$.
Even if the neutrinos have a small mass, the rate is still very small,
$\order{(m_\nu/m_W)^4}$
\cite{Petcov:1977ff,Marciano77,Lee:1977qz,Lee:1977ti}. 
However, most extensions of the standard
model, containing some new physics at the hitherto unexplored mass
scales, predict a higher rate of $\mu\to e\gamma$.  For example, in
supersymmetry (SUSY) neutrinos have heavy ``partners'', scalar sneutrinos,
whose mixing could generate  $\mu\to e\gamma$ transitions through the
interaction with charginos $\tilde\chi^\pm$, as shown in
Fig.~\ref{fig2}(a).  Scalar partners of the charged leptons,
interacting with neutralinos $\tilde\chi^0$, could also contribute to
this decay (Fig.~\ref{fig2}(b)).

Explicit supersymmetric grand unified models
\cite{Barbieri:1995tw,Barbieri:1994pv,Kuno:1999jp,Hisano:1998cx,Ellis:1999uq} 
predict a $\mu\to e\gamma$
rate just below the present 90\% CL bound from the MEGA experiment,
\cite{Brooks:1999pu},
\ba
{\Gamma(\mu\to e\gamma) \over \Gamma(\mu\to e\bar\nu_e\nu_\mu)}
< 1.2 \cdot 10^{-11}.
\ea
In the near future, a new search for $\mu\to e\gamma$ will be
undertaken at the Paul Scherrer Institute (PSI) \cite{Mori1999}, with a
single event sensitivity corresponding to the branching ratio of
$2\times 10^{-14}$.  In view of the SUSY GUT predictions, it is not
inconceivable that this experiment will find $\order{100}$ of $\mu\to
e\gamma$ decay events.  At such rate, precision studies of
lepton-number violating interactions will become possible.  It is
therefore interesting to theoretically evaluate model-independent 
electromagnetic effects which turn out to decrease the rate of
$\mu\to e\gamma$ by several percent.

\section{QED suppression of the dipole operators}
\label{sec:sup}
The effective interaction which gives rise to $\mu\to e \gamma$ has
the form 
\ba
\psfig{figure=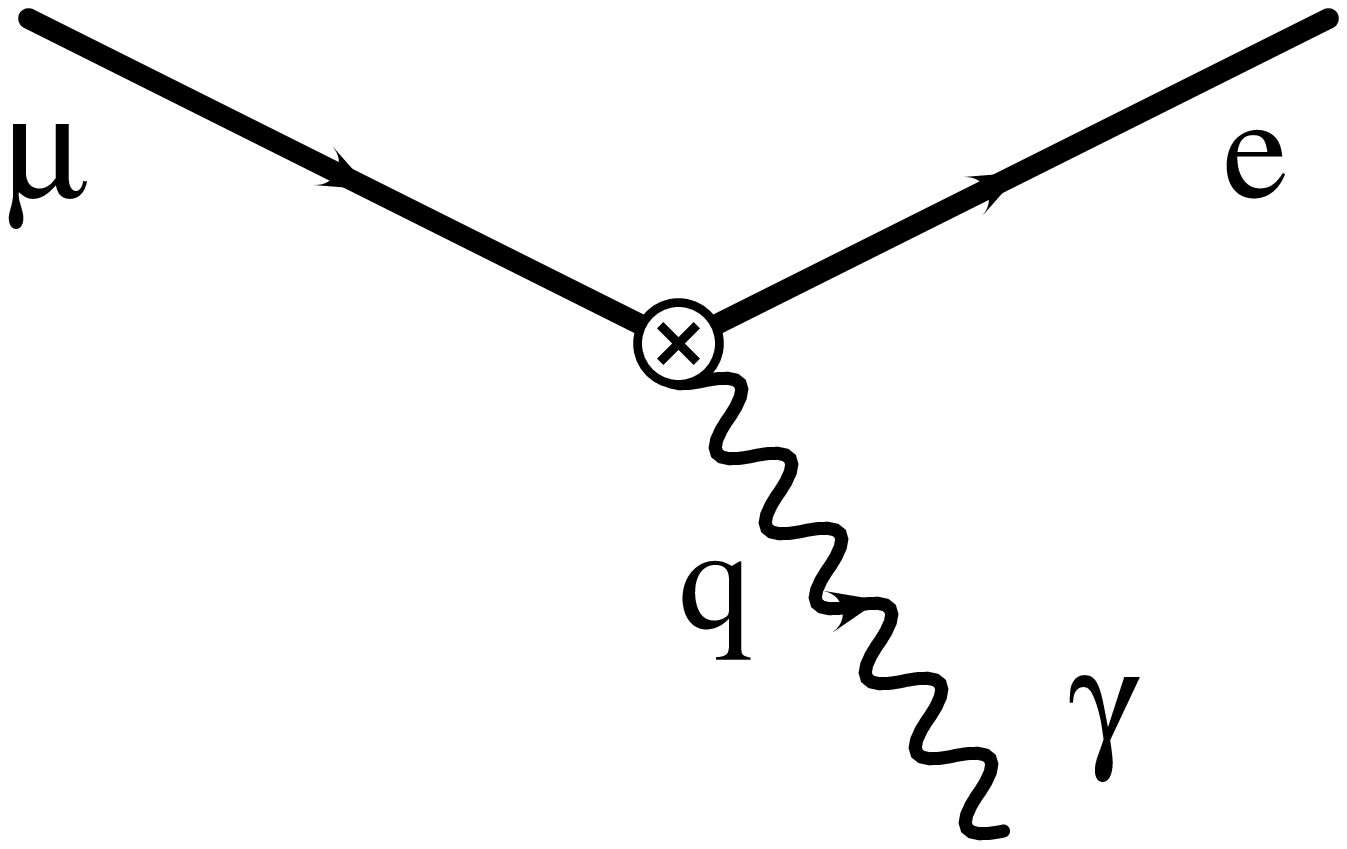,width=30mm} \raisebox{8mm}{$\;\;\;\;= \;\;
\overline{e} \; \sigma^{\mu\nu} (f_M + f_E \gamma_5)
\; \mu \cdot q_\mu A_\nu,$}
\label{eqSS}
\ea
where $f_i$ ($i=M,E$) are formfactors, calculable in explicit models
of physics beyond the standard model.  
In terms of $f_i$, the rate of
$\mu\to e\gamma$ is
\ba
\Gamma^{(0)} (\mu\to e\gamma) = {m_\mu^3\over 8\pi} 
\left(|f_M|^2 + |f_E|^2\right).
\label{eq:ra}
\ea
It is well known that the chirality-flipping electric and magnetic
dipole operators in \eq{eqSS} have (the same) large QED anomalous
dimension.  It was first computed in the context of hadron decays in
QCD
\cite{Ellis:1976uz,Wilczek:1977ry,Vainshtein:1976nd,Shifman:1978de}, 
and plays an
important role in various electromagnetic processes like the radiative
decay $b\to s\gamma$ \cite{Czarnecki:1998tn} or the muon anomalous
magnetic moment \cite{KKSS,CKM96,Degrassi:1998es} (see also
\cite{Ciuchini:1994vr}).

We denote the coefficient of the dipole-transition operators in
\eq{eqSS}, computed in a full theory violating lepton flavor, by
$f_i(\Lambda)$, where $\Lambda$ is a characteristic mass scale of the
relevant new physics.  For example, in SUSY, $f_i(\Lambda)$ would
result from the one-loop diagrams in Fig.~\ref{fig2}, and $\Lambda$
would be the characteristic mass of the superpartners.  If we now
consider an effective theory at an energy of the order of the muon
mass, the heavy exotic fields are not dynamical degrees of freedom and
we can consider the effects of Fig.~\ref{fig2} as point-like
interactions given by the Lagrangian \eq{eqSS}.

\begin{figure}[htb]
\hspace*{50mm}
\begin{minipage}{16.cm}
\vspace*{3mm}
\psfig{figure=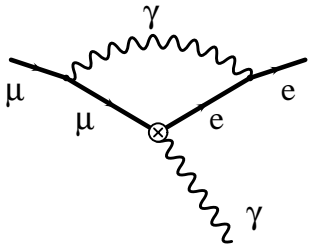,width=40mm}
\end{minipage}
\caption{\sf An example of an electromagnetic correction which
contributes to the suppression of the $\mu\to e\gamma$ decay rate.}  
\label{fig3}
\end{figure}

However, when we consider higher-order electromagnetic corrections to
this interaction, such as the one shown in Fig.~\ref{fig3}, we find
that they are logarithmically divergent in the  ultraviolet (UV).  This is
not surprising, since the dimension of the operators in \eq{eqSS} is
5, which signals non-renormalizability.  An explicit calculation shows
that the effect of those corrections amounts to 
\ba
f_i(\Lambda) \to f_i(\Lambda) \left( 1- {4\alpha\over \pi}\ln
{\Lambda\over m_\mu} + \order{\alpha} \right),
\label{qedef}
\ea 
where we have taken the UV cut-off to be equal $\Lambda$, since around
that magnitude of the loop momentum it is no longer justified to treat
the flavor-changing vertex as point-like.  The interaction is
weakened; we can denote its effective strength at the muon mass scale by
$f_i(m_\mu)$, which includes the leading logarithmic effect,
\ba
f_i(m_\mu) = f_i(\Lambda)  \left( 1- {4\alpha\over \pi}\ln
{\Lambda\over m_\mu}\right).
\ea
This effect can be quite large, since the rate \eq{eq:ra} of the decay
is proportional to the sum of squares of $f_i$,
\ba
\Gamma (\mu\to e\gamma) \simeq \left(1-{8\alpha\over
\pi}\ln{\Lambda\over m_\mu}\right)\Gamma^{(0)}(\mu\to e\gamma).
\label{rate}
\ea
If $\Lambda$ is of order 250 GeV, which is a typical SUSY mass scale
in the models considered in \cite{Barbieri:1995tw}, this corresponds
to about 14\% decrease of the rate.

It is possible to sum up the leading log effects to all orders in
$\alpha^n \ln^n\Lambda/m_\mu$ (see
e.g. \cite{Marciano:1981be,Marciano:1986pd}).  In 
the absence of mixing with other lepton-flavor non-conserving
operators, the scale dependence of the coefficients $f_i$ can be
expressed in an iterative form,
\ba
f_i(m_<) = f_i(m_>) \cdot \left( {\alpha(m_<)\over 
\alpha(m_>)} \right)^{\gamma/b}, 
\ea
where in our case the anomalous dimension is $\gamma=-8$ and $b$ is
determined using the charges $Q_j$ of all particles contributing to
the running of the fine structure constant between the scales $m_<$
and $m_>$:
\ba
b=-{4\over 3}\sum_j Q_j^2.
\ea
The explicit result for $f_i(m_\mu)$ depends on the mass spectrum of a
concrete new physics scenario.  However, higher order
leading-logarithmic effects are not expected to significantly change
the magnitude of the $\mu\to e\gamma$ rate decrease given in
\eq{rate}, because of cancelation between the running of the fine
structure constant and the effects of higher orders in the anomalous
dimension.  Similar cancelation was observed in the muon $g-2$
\cite{Degrassi:1998es}.

Typical lepton-flavor violating amplitudes, like the ones in
Fig.~\ref{fig2}, contain two new physics masses, which in general may
be quite different.  One can ask the question, what should be taken as
the argument $\Lambda$ of the logarithm in \eq{rate}.  As long as the
ratio of the two large scales is small compared to their size relative
to the muon mass, this is an issue of non-leading corrections, which
we have been neglecting.  In the case of $\mu\to
e\gamma$ induced by the small neutrino masses (where the
rate is extremely small, as discussed above), the scale $\Lambda=m_W$
in \eq{rate} is the larger of the two masses in the loop.  The inverse
of $m_W$ determines the size of the effective interaction range.

\section{Four-fermion operators}
\begin{figure}[htb]
\hspace*{5mm}
\begin{minipage}{16.cm}
\vspace*{3mm}
\begin{tabular}{ccc}
\psfig{figure=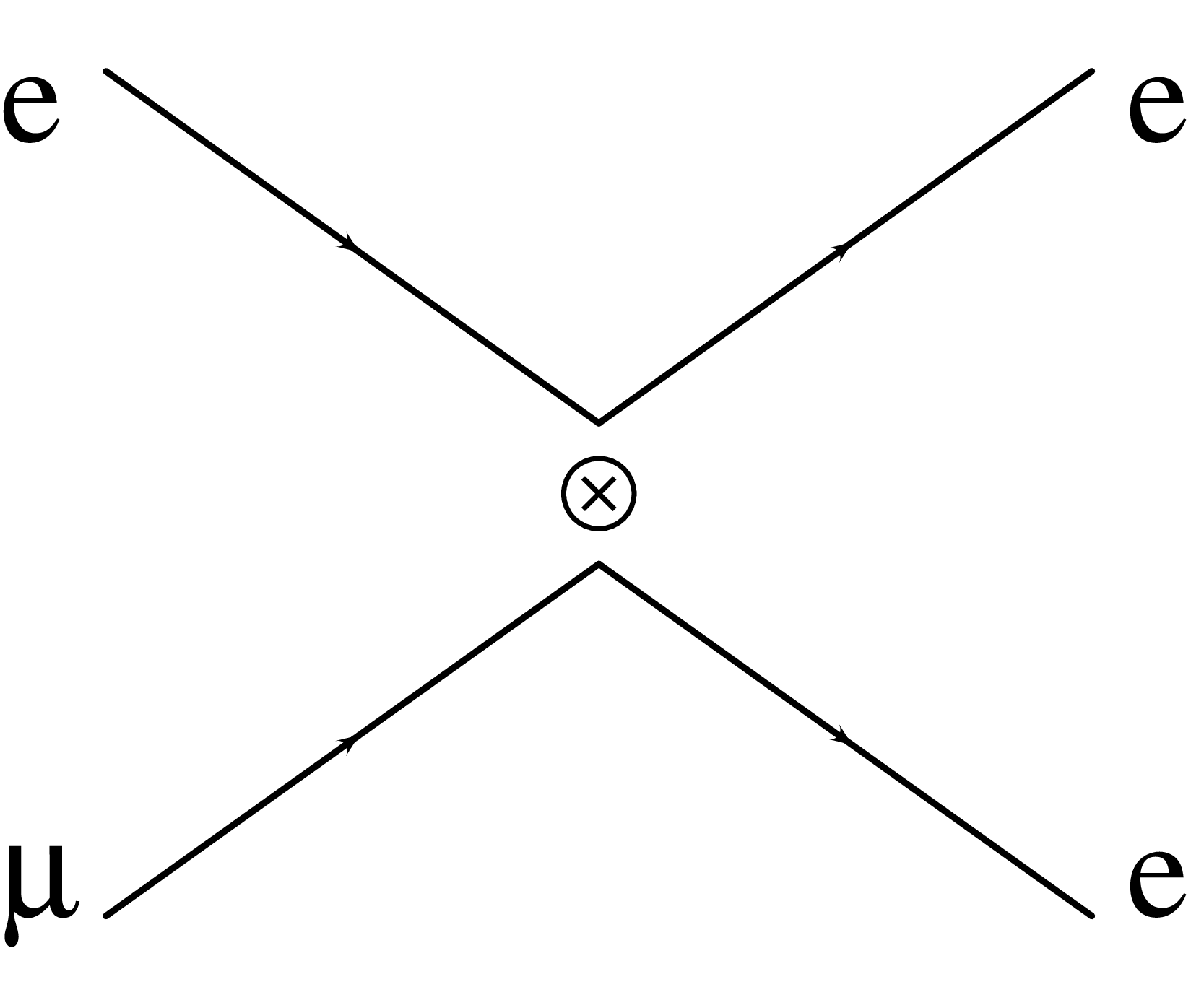,width=40mm}&
\psfig{figure=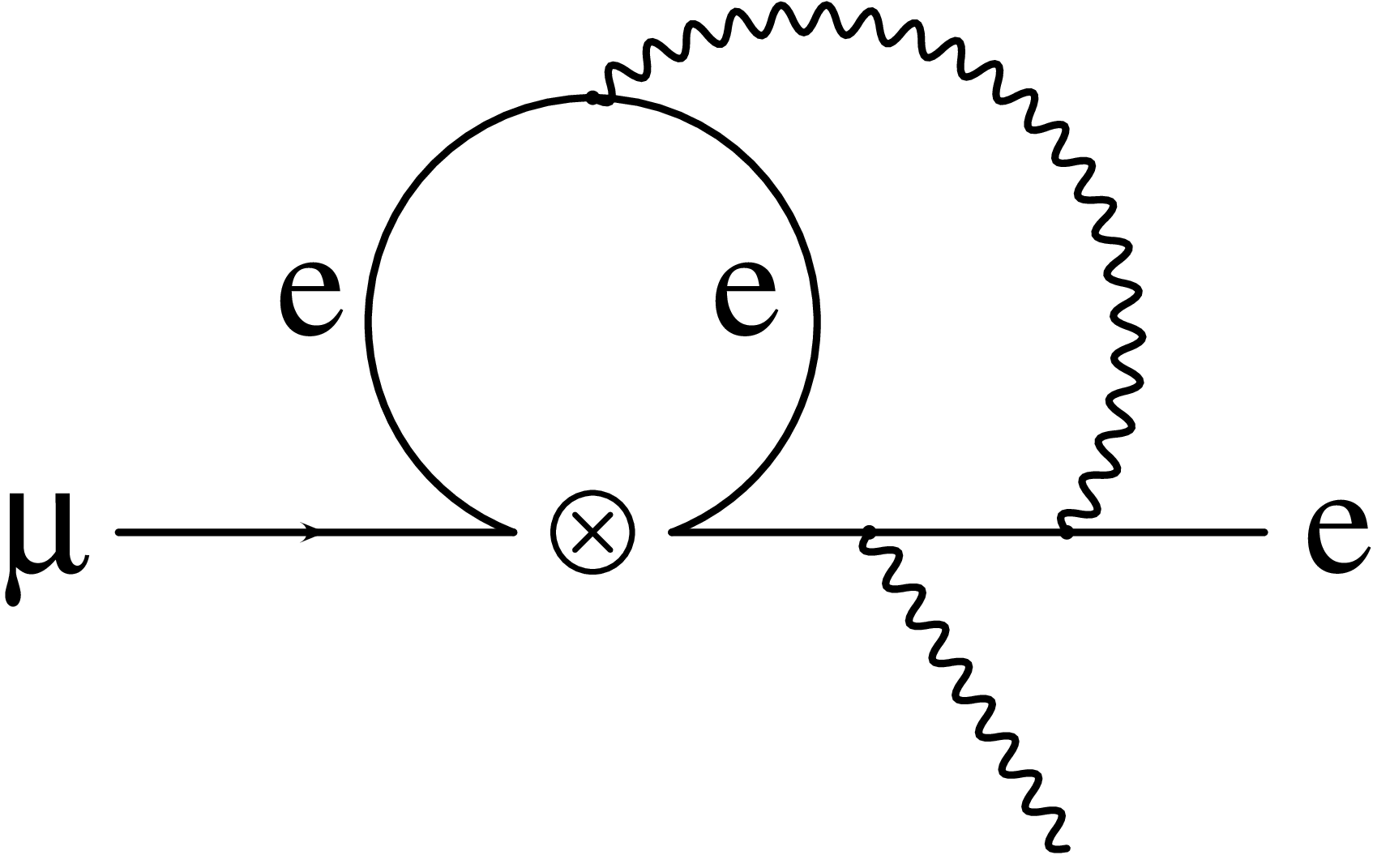,width=40mm}&
\psfig{figure=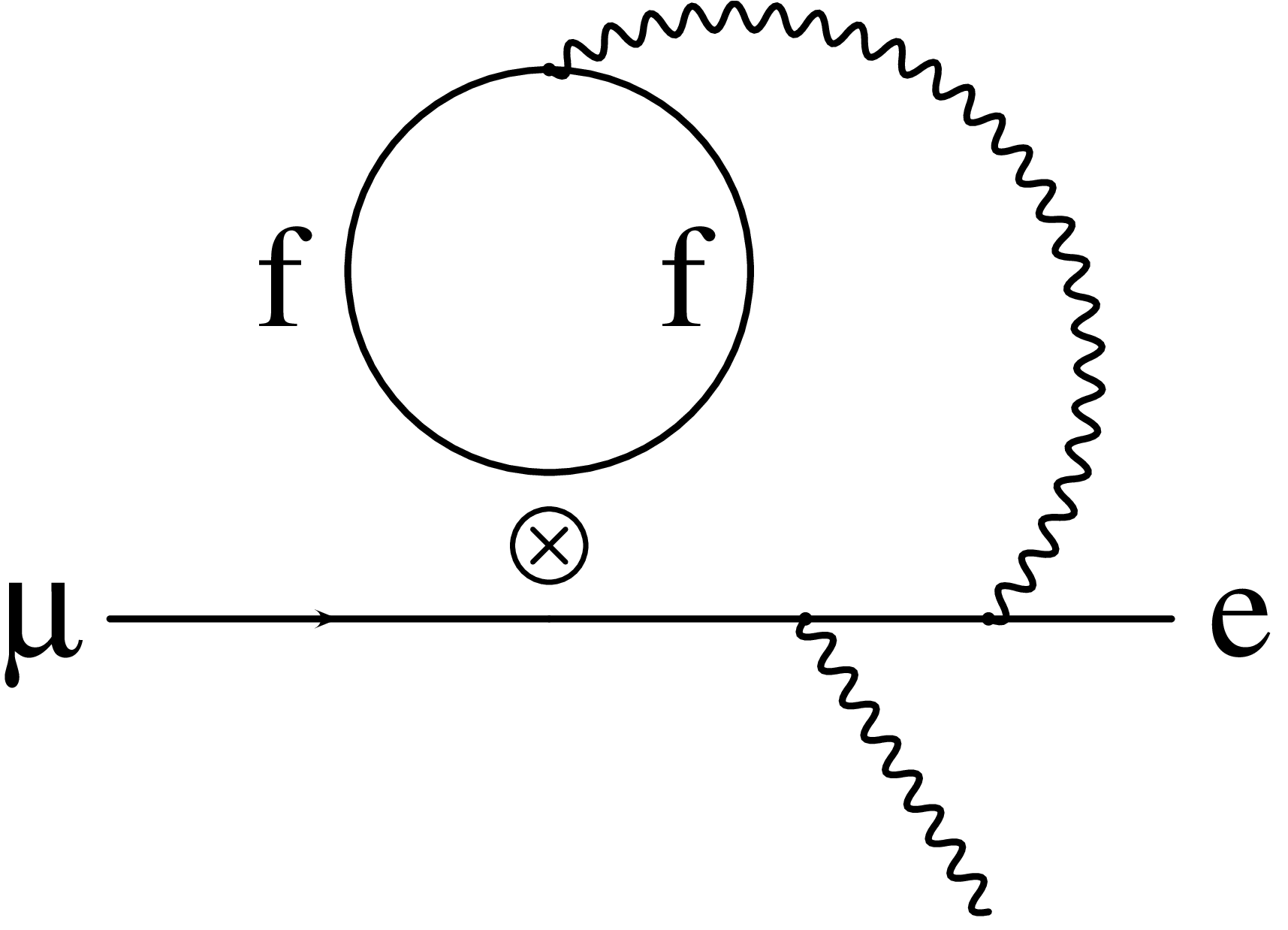,width=40mm}
\\
(a) & (b) & (c) 
\end{tabular}
\end{minipage}
\caption{\sf (a) Lepton flavor-violating four-fermion operator; (b)
Example of a contribution to $\mu\to e\gamma$ for $f=e$ or $\mu$; (c)
Example of other fermions'  contribution.}  
\label{fig4}
\end{figure}

New physics effects can also induce lepton-flavor violating
four-fermion operators such as $(\overline{e} \Gamma \mu)(\overline{f}
\Gamma f)$ (Fig.~\ref{fig4}(a)).  They contribute to $\mu\to e\gamma$
through loop effects (Fig.~\ref{fig4}(b,c)) in the same order in
${\alpha\over \pi}\ln{\Lambda\over m_\mu}$ as the suppression effect
in eq.~\eq{rate}.

In theories such as $R$-parity conserving SUSY, four-fermion
contributions are suppressed relative to the dipole operators
(Fig.~\ref{fig2}) by two powers of a coupling constant and are not
expected to contribute significantly to $\mu\to e\gamma$.  It is,
however, interesting to see to what extent we can estimate such
contributions  in a model-independent way.

Virtual fermions $f$ other than muon or electron contribute only
through ``closed'' loops, as shown in Fig.~\ref{fig4}(c).  Large
logarithms arising from such diagrams cancel at least partially in
anomaly-free theories, and we will neglect these effects.  

Here we will consider a specific example of the operator 
\ba
{\cal O}_\mathrm{x} = \mathrm{G}_\mathrm{x}
(\overline{e}\gamma^\nu L \mu)(\overline{e}\gamma_\nu L e), \qquad  
L\equiv {1-\gamma_5 \over 2},
\ea
whose anomalous dimension and mixings with other flavor-violating
operators can be found using well-known results found in studies of
the radiative quark decay $b\to s\gamma$.  
We will demonstrate that the bound on $\mathrm{G}_\mathrm{x}$
obtained from searches for $\mu\to eee$ renders the contribution of
this operator to $\mu\to e\gamma$ negligible.  We may expect that
contributions of other Dirac structures and of operators
$(\overline{e}\Gamma \mu)(\overline{\mu}\Gamma \mu)$ have similar
magnitudes.

Operator ${\cal O}_\mathrm{x}$ induces the decay  $\mu\to eee$ with a rate
\ba
\Gamma( \mu\to eee)= {\mathrm{G}_\mathrm{x}^2 m_\mu^5 \over 768\pi^3},
\ea
and we can use the bound on the branching ratio \cite{Groom:2000in},
\ba
{ \Gamma( \mu\to eee) \over  \Gamma( \mu\to e\nu \nu)} < 10^{-12},
\ea
to constrain $\mathrm{G}_\mathrm{x}$.  We find 
\ba
\mathrm{G}_\mathrm{x}
< 2\cdot 10^{-6}\mathrm{G}_\mathrm{F},
\label{boundGx}
\ea
 ($\mathrm{G}_\mathrm{F}$ is 
the Fermi constant \cite{Groom:2000in}).   

In order to find the contribution of  ${\cal O}_\mathrm{x}$ to the
amplitude $\mu\to e\gamma$ we consider its mixing with the dipole
operators in eq.~\eq{eqSS}.  We write the result as
\ba
g_\mathrm{x} \overline{e} \; \sigma^{\mu\nu} (1 + \gamma_5)
\; \mu \cdot q_\mu A_\nu,
\ea
with 
\ba
g_\mathrm{x} = {e m_\mu \mathrm{G}_\mathrm{x}\over 16\pi^2} 
{29\over 18} {\alpha\over \pi} \ln{\Lambda\over
m_\mu},
\ea
where $e=\sqrt{4\pi\alpha}\simeq 0.3$. 
Finally, we would like to compare the effect of this four-fermion
operator on the form-factors $f_i$ ($i=E,M$) with the effect of the QED
correction in eq.~\eq{qedef}.  For this purpose we assume $f_E=f_M$ and  
consider the quantity
\ba
R= {g_\mathrm{x} \over f_i {4\alpha\over \pi}\ln {\Lambda\over m_\mu}}
<
e{29 \sqrt{3} \over 8\pi \cdot 18} \cdot 10^{-6} \cdot 
{1\over  \sqrt{\mbox{BR}(\mu\to e\gamma)}},
\ea
where we have taken $f_i = {\mathrm{G}_\mathrm{F} m_\mu \over
4\sqrt{3}\pi} \sqrt{\mbox{BR}(\mu\to e\gamma)}$ and used the bound
\eq{boundGx}.  If $\mu\to e\gamma$ is discovered with a branching
ratio between $10^{-11}$ and $10^{-14}$, the upper bound on the ratio
$R$ of the four-fermion and dipole radiative effects  will be between about
$10^{-2}$ and 0.3.  

The QED corrections we considered in this paper will be relevant for
the upcoming PSI experiment if it observes a fair number (of the order
of a hundred or more) of decay events $\mu\to e\gamma$.  This
corresponds to the branching ratio of at least $10^{-12}$, for which
the ratio $R$ is about 0.03.  We conclude that the effects of the
four-fermion operators are likely to be negligible for the next
generation of the $\mu\to e\gamma$ searches.

\section{Conclusions}

The logarithmic suppression which we have discussed in Section
\ref{sec:sup} 
affects not only $\mu\to e\gamma$ but also other 
lepton-flavor violating processes occurring via the dipole transition
of the type \eq{eqSS}.  For example, the rates of the $\tau$-lepton
decays $\tau\to \mu\gamma$ and  $\tau\to e\gamma$ are decreased by
\ba
1-{8\alpha\over \pi} \ln {\Lambda\over m_\tau},
\ea
which is about $7.5\ldots 12$\% for $\Lambda=100\ldots 1000$ GeV.  On
the other hand, the decays of the type $\mu^+ \to e^+e^+e^-$ and
muon-electron conversion in the nuclear field, $\mu^- N\to e^- N$, can
occur via a more general interaction, including monopole formfactors,
which do not receive such logarithmic corrections.

To summarize, we have pointed out an electromagnetic short-distance
effect which decreases the predicted rate of the lepton-flavor
violating decay $\mu \to e\gamma$ by a factor $\left(
1-{8\alpha\over\pi} \ln {\Lambda\over m_\mu}\right)$, or $12\ldots
17$\% for the new physics scale $\Lambda =100\ldots 1000$ GeV.  If the
lepton-flavor non-conservation is observed by the next generation of
experiments, the $\mu \to e\gamma$ search at the PSI and the
conversion $\mu^- N\to e^- N$ search MECO in Brookhaven, this
correction will help disentangle the underlying new physics structure.

\section*{Acknowledgments}
We thank John Ellis and William Marciano for helpful discussions.  AC
thanks the Brookhaven Laboratory High Energy Theory Group for
hospitality during the work on this problem.  This research was
supported in part by the Natural Sciences and Engineering Research
Council of Canada.


\end{document}